\documentclass[prl,10pt,showpacs,amsmath,twocolumn,floatfix]{revtex4}
\usepackage{amsmath}
\usepackage{amsfonts}
\usepackage{graphicx}

\newcommand{\parti}[2]{\frac{\partial #1}{\partial #2}}
\newcommand{\partit}[2]{\frac{\partial^2 #1}{\partial #2^2}}

\newcommand{\avg}[1]{\langle#1\rangle}

\newcommand{\bs}[1]{#1}
\newcommand{\boldsym}[1]{#1}

\newcommand{\bk}[1]{\left(#1\right)}
\newcommand{\Bk}[1]{\left[#1\right]}
\newcommand{\BK}[1]{\left\{#1\right\}}

\newcommand{\trace}{\operatorname{tr}}
\newcommand{\tfinal}{T}
\begin{document}
\title{Time-Symmetric Quantum Theory of Smoothing}

\author{Mankei Tsang}

\email{mankei@mit.edu}

\affiliation{Research Laboratory of Electronics,
Massachusetts Institute of Technology, Cambridge, Massachusetts
02139, USA}






\date{\today}

\begin{abstract}
  Smoothing is an estimation technique that takes into account both
  past and future observations, and can be more accurate than
  filtering alone. In this Letter, a quantum theory of smoothing is
  constructed using a time-symmetric formalism, thereby generalizing
  prior work on classical and quantum filtering, retrodiction, and
  smoothing. The proposed theory solves the important problem of
  optimally estimating classical Markov processes coupled to a quantum
  system under continuous measurements, and is thus expected to find
  major applications in future quantum sensing systems, such as
  gravitational wave detectors and atomic magnetometers.
\end{abstract}

\maketitle
\noindent
Estimation theory is concerned with the inference of unknown signals,
given their \textit{a priori} statistics as well as noisy observations
\cite{jazwinski}. Depending on the time at which the signal is to be
estimated relative to the observation time interval, estimation
problems can be divided into four classes: \textit{Prediction}, the
estimation of a signal at time $\tau$ given observations before
$\tau$; \textit{Filtering}, given observations before and up to
$\tau$; \textit{Smoothing}, given observations before and after
$\tau$; and \textit{Retrodiction}, given observations after $\tau$
\cite{retro}.  Among the four classes, prediction and filtering have
received the most attention, given their importance in applications
that require real-time knowledge of a system, such as control, weather
forecast, and quantitative finance.  If we allow delay in the
estimation, however, we can take into account the more advanced
observations to produce a more accurate estimation of the signal some
time in the past via smoothing techniques.  For this reason, smoothing
is mainly used in communication and sensing applications, when
accuracy is paramount but real-time data are not required.

Conventional quantum theory can be regarded as a prediction
theory. The quantum state in the Schr\"odinger picture represents our
maximal knowledge of a system given prior observations. In particular,
the quantum filtering theory developed by Belavkin and others
\cite{belavkin,carmichael} can be regarded as a generalization of the
classical nonlinear filtering theory devised by Stratonovich and
Kushner \cite{kushner}. Quantum smoothing and retrodiction theories,
on the other hand, have been proposed by Aharonov \textit{et al.}\ as
an alternative formulation of quantum mechanics \cite{aav}, Barnett
\textit{et al.}\ for the purpose of parameter estimation
\cite{barnett}, and Yanagisawa for initial quantum state estimation
\cite{yanagisawa}.  In this Letter, I generalize these earlier results
on classical and quantum estimation to a quantum theory of smoothing
for continuous waveform estimation. I am primarily interested in the
estimation of classical random processes, such as gravitational waves
and magnetic fields, coupled to a quantum object, such as a quantum
mechanical oscillator or an atomic spin ensemble, under continuous
measurements. Previous studies on the use of filtering for these
estimation problems \cite{mabuchi} model the classical signals in
terms of constant parameters or waveforms with deterministic
evolution, but it is more desirable to model them as Markov processes
for generality and robustness, in which case smoothing can be
significantly more accurate than filtering \cite{jazwinski}. Quantum
estimation of a random optical phase process has recently been studied
by Wiseman and co-workers \cite{wiseman_phase,berry} and Tsang
\textit{et al.}\ \cite{tsang}, but a general quantum smoothing theory
is still lacking. The theory proposed here is thus expected to find
important applications in future quantum sensing systems, such as
gravitational wave detectors and atomic magnetometers.


\begin{figure}[htbp]
\includegraphics[width=0.45\textwidth]{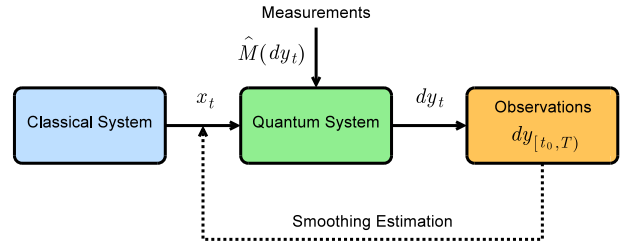}
\caption{(Color online). Schematic of the continuous waveform
  estimation problem.}
\label{scheme}
\end{figure}

Consider the estimation problem schematically shown in
Fig.~\ref{scheme}.  A vectoral classical random process $\bs x_t
\equiv [x_1(t),\dots,x_n(t)]^\textrm{T}$ is coupled to a quantum
system. The back-action of the quantum system on the classical system
that produces $\bs x_t$ is assumed to be negligible, so that the
statistics of $\bs x_t$ remain unperturbed and classical.  This
assumption should be satisfied for the purpose of sensing and avoids
the contentious issue of quantum back-action on classical systems
\cite{backaction}.  The quantum system is measured continuously, via a
weak measurement operator $\hat M(d\bs y_t)$, where $d\bs y_t \equiv
[dy_1(t),\dots,dy_m(t)]^\textrm{T}$ is the vectoral measurement
outcome at time $t$. Define the observations in the time interval
$[t_1,t_2)$ as $d\bs y_{[t_1,t_2)} \equiv \BK{d\bs y_t, t_1\le t <
  t_2}$.  My ultimate goal is to calculate the fixed-interval
smoothing probability density $P(\bs x_\tau|d\bs y_{[t_0,\tfinal)})$
at time $\tau$, conditioned upon past and future observations in the
time interval $t_0\le \tau \le \tfinal$, so that the conditional
expectations of $\bs x_\tau$ and the associated errors can be
determined.

Central to my derivation is the use of a hybrid classical-quantum
density operator $\hat\rho_t(\bs x_t)$, which provides joint classical
and quantum statistics at time $t$ \cite{backaction,warszawski}.  The classical
probability density for $\bs x_t$ and the unconditional density
operator can be determined from the hybrid operator by
\begin{align}
P(\bs x_t) &= \trace\Bk{\hat\rho_t(\bs x_t)},&
\hat\rho_t &= \int d\bs x_t\hat\rho_t(\bs x_t),
\label{uncond}
\end{align}
respectively. To derive the smoothing density, I will need the
conditional hybrid density operator $\hat\rho_\tau(\bs x_\tau|d\bs
y_{[t_0,\tau)})$ given past observations, and also a hybrid effect
operator, $\hat E_\tau(d\bs y_{[\tau,\tfinal)}|\bs x_\tau)$, which
determines the joint statistics of future observations $d\bs
y_{[\tau,\tfinal)}$ given an arbitrary hybrid density operator
$\hat\rho_\tau(\bs x_\tau)$ at time $\tau$,
\begin{align}
P\bk{d\bs y_{[\tau,\tfinal)}|\hat\rho_\tau(\bs x_\tau)}
&= \int d\bs x_\tau \trace\Bk{\hat E_\tau(d\bs y_{[\tau,\tfinal)}|\bs x_\tau)
\hat\rho_\tau(\bs x_\tau)}.
\label{trace}
\end{align}
The smoothing probability density is then
\begin{align}
&\quad P(\bs x_\tau|d\bs y_{[t_0,\tfinal)}) 
\nonumber\\
&= P(\bs x_\tau|d\bs y_{[t_0,\tau)}, d\bs y_{[\tau,\tfinal)})
=\frac{P(\bs x_\tau,d\bs y_{[\tau,\tfinal)}|d\bs y_{[t_0,\tau)})}
{P(d\bs y_{[\tau,\tfinal)}|d\bs y_{[t_0,\tau)})}
\nonumber\\
&= \frac{\trace[\hat E_\tau(d\bs y_{[\tau,\tfinal)}|\bs x_\tau)
\hat\rho_\tau(\bs x_\tau|d\bs y_{[t_0,\tau)})]}
{\int d\bs x_\tau\trace[\hat E_\tau(d\bs y_{[\tau,\tfinal)}|\bs x_\tau)
\hat\rho_\tau(\bs x_\tau|d\bs y_{[t_0,\tau)})]}.
\end{align}
To calculate the conditional hybrid density operator
$\hat\rho_\tau(\bs x_\tau|d\bs y_{[t_0,\tau)})$, which also solves the
filtering problem, first consider the conditional density operator
$\hat\rho_\tau(|\bs x_{[t_0,\tau)})$ in discrete time, which describes
the quantum state given a particular trajectory of $\bs x_{[t_0,\tau)}
\equiv \{\bs x_{t_0},\bs x_{t_0+\delta t},\dots,\bs x_{\tau-\delta
  t}\}$,
\begin{align}
\hat\rho_\tau(|\bs x_{[t_0,\tau)})
&= \mathcal K(\bs x_{\tau-\delta t})
\dots \mathcal K(\bs x_{t_0+\delta t})\mathcal K(\bs x_{t_0})
\hat\rho_{t_0},
\label{traj}
\end{align}
where $\hat\rho_{t_0}$ is the initial \textit{a priori} density
operator, $\mathcal K(\bs x_t)\equiv \exp[\delta t\mathcal L(\bs
x_t)]$ is a super-operator that governs the quantum system evolution
for the time interval $\delta t$ independent of the measurement
process, $\mathcal L$ is a super-operator in Lindblad form, and $\bs
x_t$ acts as a parameter of the evolution.
Averaging over trajectories of $\bs x_{[t_0,\tau)}$, the hybrid
density operator $\hat\rho_\tau(\bs x_\tau)$ can be expressed as
\begin{align}
\hat\rho_\tau(\bs x_\tau)
&= \int d\bs x_{\tau-\delta t} \dots d\bs x_{t_0}
\hat\rho_\tau(|\bs x_{[t_0,\tau)})P(\bs x_{[t_0,\tau)},\bs x_\tau).
\label{hybrid_avg}
\end{align}
This expression can be verified by substituting it into
Eqs.~(\ref{uncond}). If $\bs x_t$ is a Markov process, $P(\bs
x_{[t_0,\tau)},\bs x_\tau)=P(\bs x_{[t_0,\tau]}) = P(\bs x_\tau|\bs
x_{\tau-\delta t})\dots P(\bs x_{t_0+\delta t}|\bs x_{t_0})P(\bs
x_{t_0})$, $P(\bs x_{t_0})$ being the initial \textit{a priori}
probability density.  Rearranging the terms in Eqs.~(\ref{traj}) and
(\ref{hybrid_avg}), $\hat\rho_\tau(\bs x_\tau)$ can be solved by
iterating the formula
\begin{align}
\hat\rho_{t+\delta t}(\bs x_{t+\delta t})
&= \int d\bs x_t P(\bs x_{t+\delta t}|\bs x_t)
\mathcal K(\bs x_t)
\hat\rho_{t}(\bs x_{t}),
\label{iterate}
\end{align}
with the initial condition $\hat\rho_{t_0}(\bs x_{t_0}) =
\hat\rho_{t_0}P(\bs x_{t_0})$. $P(\bs x_{t+\delta t}|\bs x_t)$ for an important
class of Markov processes can be determined from the It\=o stochastic
differential equation \cite{jazwinski}
\begin{align}
d \bs x_t = \bs A(\bs x_t,t) dt+ \bs B(\bs x_t,t) d\bs W_t,
\label{ito}
\end{align}
where $d\bs W_t$ is a vectoral Wiener increment with $\mathcal E\{d
\bs W_t\} = 0$ and $\mathcal E\{d\bs W_t d\bs W_t^\textrm{T}\} \equiv \bs Q(t)
dt$.

To calculate the \textit{a posteriori} hybrid state after a
measurement, the quantum Bayes theorem \cite{carmichael} can be
generalized as
\begin{align}
\hat\rho_{t}(\bs x_{t}|\delta\bs y_t)
&= \frac{\mathcal J(\delta\bs y_t)
\hat\rho_{t}(\bs x_{t})}
{\int dx_{t}\trace[\mathcal J(\delta\bs y_t)
\hat\rho_{t}(\bs x_{t})]},
\end{align}
where $\mathcal J(\delta\bs y_t)\hat\rho \equiv \hat M(\delta\bs
y_t)\hat\rho \hat M^\dagger(\delta\bs y_t)$. The evolution of the
hybrid density operator conditioned upon past observations $\delta
y_{[t_0,t)}\equiv \{\delta y_{t_0},\delta y_{t_0+\delta t},
\dots,\delta y_{t-\delta t}\}$ is therefore given by
\begin{align}
&\quad
\hat\rho_{t+\delta t}(\bs x_{t+\delta t}|\delta\bs y_{[t_0,t)},\delta\bs y_t)
\nonumber\\
&= 
\frac{\int d\bs x_t P(\bs x_{t+\delta t}|\bs x_t)
\mathcal K(\bs x_t)\mathcal J(\delta\bs y_t)
\hat\rho_{t}(\bs x_{t}|\delta\bs y_{[t_0,t)})}
{\int dx_{t}\trace[\mathcal J(\delta\bs y_t)
\hat\rho_{t}(\bs x_{t}|\delta\bs y_{[t_0,t)})]}.
\end{align}
Assuming Gaussian measurements, the measurement operator in the
continuous limit is \cite{belavkin,carmichael,diosi}
\begin{align}
\hat M(dz_t) &\propto
\hat 1+\sum_{\mu}\gamma_\mu(t)
\Bk{\frac{1}{2}(dz_t)_\mu\hat{c}_\mu
- \frac{d t}{8}\hat{c}_\mu^\dagger\hat{c}_\mu},
\end{align}
where $\gamma_\mu$ is assumed to be positive, $dz_t$ is a vectoral
observation process, and $\hat c$ is a vector of arbitrary
operators. Defining $dy_t \equiv Udz_t$ and $\hat C \equiv U\hat c$,
$U$ being a unitary matrix, the measurement operator can be cast into
an equivalent but slightly more useful form as
\begin{align}
\hat M(d\bs y_t) &\propto
\hat 1+\frac{1}{2}d\bs y_t^\textrm{T}\bs R^{-1}(t)\hat{\bs C}
- \frac{d t}{8}\hat{\bs C}^{\dagger \textrm{T}}\bs R^{-1}(t)\hat{\bs C},
\label{measure}
\end{align}
where $\bs R$ is a real positive-definite matrix with eigenvalues
$1/\gamma_\mu$. The stochastic master equation for $\hat\rho_t(\bs
x_t=\bs x|d\bs y_{[t_0,t)}) \equiv \hat F(\bs x,t)$ in the It\=o sense
is hence
\begin{widetext}
\begin{align}
d\hat F
&= dt\bigg\{\mathcal L(\bs x)\hat F
-\sum_\mu \parti{}{x_\mu}
\bk{A_\mu \hat F}
+\frac{1}{2}\sum_{\mu,\nu}
\parti{^2}{x_\mu\partial x_\nu}
\Bk{\bk{\bs B\bs Q\bs B^\textrm{T}}_{\mu\nu}\hat F}
\nonumber\\&\quad
+\frac{1}{8}\bk{2\hat{\bs C}^\textrm{T}\bs R^{-1}\hat F\hat{\bs C}^\dagger-
\hat{\bs C}^{\dagger \textrm{T}} \bs R^{-1}\hat{\bs C}\hat F
-\hat F\hat{\bs C}^{\dagger \textrm{T}} \bs R^{-1}\hat{\bs C}
}\bigg\}
+\frac{1}{2}\Bk{d\boldsym\eta_t^\textrm{T}\bs R^{-1}
\bk{\hat{\bs C}-\avg{\hat{\bs C}}_{\hat F}}\hat F+
\textrm{H.c.}},
\label{hybrid_kushner}
\end{align}
where $d\boldsym\eta_t \equiv d\bs y_t- dt\avg{\hat{\bs C}+\hat{\bs
    C}^\dagger}_{\hat F}/2$ is a real vectoral Wiener increment with
covariance matrix $Rdt$, $\avg{\hat{\bs C}}_{\hat F} \equiv \int d\bs
x \trace[\hat{\bs C} \hat F(\bs x,t)]$, and H.c.\ denotes the
Hermitian conjugate.  Equation (\ref{hybrid_kushner}) solves the
filtering problem for the hybrid classical-quantum system and
generalizes the Kushner equation \cite{jazwinski,kushner} and the
Belavkin equation \cite{belavkin}.  The continuous phase estimation
theory proposed in Ref.~\cite{berry} may be considered as a special
case of Eq.~(\ref{hybrid_kushner}).  A linear version of the master
equation for an unnormalized $\hat F(\bs x,t)$, analogous to the
classical Zakai equation \cite{zakai}, is
\begin{align}
d\hat f
&= dt\bigg\{\mathcal L(\bs x)\hat f
- \sum_\mu\parti{}{x_\mu}
\bk{A_\mu \hat f}
+\frac{1}{2}\sum_{\mu,\nu}
\parti{^2}{x_\mu\partial x_\nu}
\Bk{\bk{\bs B\bs Q\bs B^\textrm{T}}_{\mu\nu}\hat f}
\nonumber\\&\quad
+\frac{1}{8}\bk{2\hat{\bs C}^\textrm{T}\bs R^{-1}\hat f\hat{\bs C}^\dagger-
\hat{\bs C}^{\dagger\textrm{T}}\bs R^{-1}\hat{\bs C}\hat f-\hat f\hat{\bs C}^{\dagger\textrm{T}}\bs R^{-1} \hat{\bs C}}\bigg\}
+\frac{1}{2}\bk{d\bs y_t^\textrm{T}\bs R^{-1} \hat{\bs C}\hat f+ \textrm{H.c.}},
\label{hybrid_zakai}
\end{align}
and $\hat F(\bs x,t)$ is given by $\hat f(\bs x,t)/\int d\bs
x\trace[\hat f(\bs x,t)]$.

To solve for $\hat E_\tau(d\bs y_{[\tau,\tfinal)}|\bs x_\tau)$, rewrite
Eq.~(\ref{trace}) in discrete time as
\begin{align}
P(\delta\bs y_{[\tau,\tfinal)}|\hat\rho_\tau(\bs x_\tau))
&=\int d\bs x_\tau \trace\Bk{\hat E_\tau(\delta\bs y_{[\tau,\tfinal)}|\bs x_\tau)
\hat\rho_\tau(\bs x_\tau)}
\label{compare1}\\
&=\int d\bs x_{\tfinal}\trace\bigg[\int d\bs x_{\tfinal-\delta t}P(\bs x_{\tfinal}|\bs x_{\tfinal-\delta t})
\mathcal K(\bs x_{\tfinal-\delta t})\mathcal J(\delta\bs y_{\tfinal-\delta t})
\dots \int d\bs x_\tau P(\bs x_{\tau+\delta t}|\bs x_\tau)
\mathcal K(\bs x_\tau)\mathcal J(\delta\bs y_\tau)
\hat\rho_\tau(\bs x_\tau)\bigg].
\label{compare2}
\end{align}
Comparing Eq.~(\ref{compare1}) with Eq.~(\ref{compare2}) and defining
the adjoint of a super-operator $\mathcal O$ as $\mathcal O^*$, such
that $\trace[\hat E (\mathcal O\hat \rho)] = \trace[(\mathcal O^*\hat
E)\hat\rho]$, the hybrid effect operator can be expressed as
\begin{align}
\hat E_\tau(\delta\bs y_{[\tau,\tfinal)}|\bs x_\tau)
&= 
\mathcal J^*(\delta\bs y_\tau)
\mathcal K^*(\bs x_\tau)
\int d\bs x_{\tau+\delta t}P(\bs x_{\tau+\delta t}|\bs x_\tau)
\dots
\mathcal J^*(\delta\bs y_{\tfinal-\delta t})
\mathcal K^*(\bs x_{\tfinal-\delta t})
\int d\bs x_{\tfinal} P(\bs x_{\tfinal}|\bs x_{\tfinal-\delta t})\hat 1.
\end{align}
The stochastic master equation for an unnormalized $\hat E_t(d\bs
y_{[t,\tfinal)}|\bs x_t=\bs x)\propto \hat g(\bs x,t)$ in continuous time
becomes
\begin{align}
-d\hat g &= dt\bigg[\mathcal L^*(\bs x)\hat g
+\sum_\mu A_\mu \parti{}{x_\mu}
\hat g
+\frac{1}{2}\sum_{\mu,\nu}\bk{\bs B\bs Q\bs B^\textrm{T}}_{\mu\nu}
\parti{^2}{x_\mu\partial x_\nu}\hat g
\nonumber\\&\quad
+\frac{1}{8}\bk{2\hat{\bs C}^{\dagger\textrm{T}}\hat g\bs R^{-1}\hat{\bs C}
-\hat g\hat{\bs C}^{\dagger\textrm{T}}\bs R^{-1}\hat{\bs C}-
\hat{\bs C}^{\dagger\textrm{T}}\bs R^{-1}\hat{\bs C}\hat g}\bigg]
+\frac{1}{2}\bk{d\bs y_t^\textrm{T}\bs R^{-1}
\hat g\hat{\bs C} + \textrm{H.c.}},
\label{retro_zakai}
\end{align}
\end{widetext}
which is the adjoint equation of Eq.~(\ref{hybrid_zakai}), to be
solved backward in time using the backward It\=o rule and the final
condition $\hat g(\bs x,\tfinal)\propto \hat 1$.  The smoothing probability
density is hence
\begin{align}
  h(\bs x,\tau)\equiv P(\bs x_\tau=\bs x|d\bs y_{[t_0,\tfinal)}) =
  \frac{\trace[\hat g(\bs x,\tau)\hat f(\bs x,\tau)]} {\int d\bs x
    \trace[\hat g(\bs x,\tau)\hat f(\bs x,\tau)]}.
\label{smooth}
\end{align}
This form of smoothing, which combines the solutions of adjoint
equations (\ref{hybrid_zakai}) and (\ref{retro_zakai}), has a pleasing
time symmetry, and can be regarded as a generalization of the
classical nonlinear two-filter smoothing theory proposed by Pardoux
\cite{pardoux}.

Equations (\ref{hybrid_kushner}), (\ref{hybrid_zakai}),
(\ref{retro_zakai}), and (\ref{smooth}) are the central results of
this Letter and form the basis of a general quantum prediction,
filtering, smoothing, and retrodiction theory for continuous waveform
estimation. One way of solving them is to convert them to stochastic
partial differential equations for quasi-probability
distributions. For quantum systems with continuous degrees of freedom,
the Wigner distribution is especially helpful. Let $f(\bs q,\bs p,\bs
x,t)$ and $g(\bs q,\bs p,\bs x,t)$ be the Wigner distributions of
$\hat f(\bs x,t)$ and $\hat g(\bs x,t)$, respectively. They have the
desired property $\int d\bs qd\bs p\, g(\bs q,\bs p,\bs x,t)f(\bs
q,\bs p,\bs x,t) \propto\trace[\hat g(\bs x,t)\hat f(\bs x,t)]$, which
is unique among generalized quasi-probability distributions
\cite{mandel}. The smoothing density can then be rewritten as
\begin{align}
h(\bs x,\tau)
&= \frac{\int d\bs qd\bs p\, g(\bs q,\bs p,\bs x,\tau)
f(\bs q,\bs p,\bs x,\tau)}
{\int d\bs x d\bs qd\bs p\, g(\bs q,\bs p,\bs x,\tau)
f(\bs q,\bs p,\bs x,\tau)}.
\label{wigner_smooth}
\end{align}
As an illustration of the smoothing theory, consider the estimation of
a classical force, say $x_1(t)$, acting on a quantum mechanical
harmonic oscillator, and the position of the oscillator is monitored,
via an optical phase-locked loop for example
\cite{wiseman_phase,berry,tsang}. Let $\mathcal L\hat\rho = -\mathcal
L^*\hat\rho= -(i/\hbar)[\hat H,\hat\rho]$, $\hat H = (\hat p^2 +
\omega^2 \hat q^2)/2- x_1\hat q$, and $\hat{\bs C} = \hat q$.  The
linear stochastic equations for the Wigner distributions become
\begin{align}
df
&= dt\bigg\{- p\parti{f}{q}+\bk{\omega^2 q- x_1}\parti{f}{p}
- \sum_\mu\parti{}{x_\mu}
\bk{A_\mu f}
\nonumber\\&\quad
+\frac{1}{2}
\sum_{\mu,\nu}
\parti{^2}{x_\mu\partial x_\nu}
\Bk{\bk{\bs B\bs Q\bs B^\textrm{T}}_{\mu\nu}f}
+\frac{\hbar^2}{8R} \partit{f}{p}\bigg\}
\nonumber\\&\quad
+\frac{dy_t  q}{R}  f,
\end{align}
and
\begin{align}
-dg &= dt\bigg[p\parti{g}{q}-
\bk{\omega^2 q- x_1}\parti{g}{p}
+ \sum_\mu A_\mu\parti{g}{x_\mu}
\nonumber\\&\quad
+\frac{1}{2}\sum_{\mu,\nu}\bk{\bs B\bs Q\bs B^\textrm{T}}_{\mu\nu}
\parti{^2g}{x_\mu\partial x_\nu}
+\frac{\hbar^2}{8R} \partit{g}{p}
\bigg]
+\frac{dy_t q}{R} g.
\end{align}
These equations are then identical to the classical forward and
backward Zakai equations \cite{zakai,pardoux}. If $\bs x_t$ is
Gaussian and the initial $f$ is Gaussian, the means and covariances of
the Gaussian $f$, $g$, and $h$ can be obtained using the
Mayne-Fraser-Potter two-filter smoother \cite{mayne,tsang}, which
calculates those of $f$ and $g$ using forward and backward Kalman-Bucy
filters \cite{jazwinski}, and then combines them to give the means and
covariances of $h$. As is well known in classical estimation theory,
unless $x_1$ is constant, the smoothing estimates and covariances
cannot be obtained from a filtering theory alone. The reduced
estimation errors associated with quantum smoothing can in principle
be verified experimentally in future quantum sensing systems.

Discussions with Seth Lloyd and Jeffrey Shapiro are gratefully
acknowledged. This work is supported by the Keck Foundation Center for
Extreme Quantum Information Theory.


\begin{thebibliography}{99}
\bibitem{retro}In classical estimation theory terminology,
  retrodiction is the same as smoothing, but here we use the former
  term to define another class of estimation problems, in order to
  better classify prior work in quantum estimation theory.
\end{thebibliography}
\end{document}